\documentclass[nofootinbib,twocolumn,superscriptaddress,showpacs,preprintnumbers,amsmath,amssymb,aps,prd]{revtex4}
\usepackage{amssymb} \usepackage{graphicx} \usepackage{color} \usepackage{hyperref} \usepackage[mathscr]{euscript}

  \def\mwm#1{matter-without-matter#1
  (MWM#1)\gdef\mwm{MWM}}
\def\bh#1{black hole#1
  (BH#1)\gdef\bh{BH}}
  \def\ahz#1{apparent horizon#1 (AH#1)\gdef\ahz{AH}}
  \def\bbh#1{binary black hole#1
  (BBH#1)\gdef\bbh{BBH}}
  
\begin{document}

\title{Super-Extremal Spinning Black Holes via Accretion}

\author{Tanja Bode}
\affiliation{Center for Relativistic Astrophysics and
School of Physics\\
Georgia Institute of Technology, Atlanta, Georgia 30332, USA}
\author{Pablo Laguna}
\affiliation{Center for Relativistic Astrophysics and
School of Physics\\
Georgia Institute of Technology, Atlanta, Georgia 30332, USA}
\author{Richard Matzner}
\affiliation{Center for Relativity and Department of Physics\\
University of Texas at Austin,
Austin, Texas 78712, USA}

\begin{abstract}   

A Kerr black hole with mass $M$ and angular momentum $J$ satisfies 
the extremality inequality $|J| \le M^2$. In the presence of matter and/or 
gravitational radiation, this bound needs to be reformulated in terms of local 
measurements of the mass and the angular momentum directly associated with the black hole. The isolated 
and dynamical horizon framework provides such quasi-local characterization 
of black hole mass and angular momentum. With this framework, it is possible in axisymmetry to reformulate the extremality limit 
as $|J| \le 2\,M_H^2$, with $M_H$ the irreducible mass of the black hole computed from its apparent 
horizon area and $J$ obtained using approximate rotational Killing vectors on the 
apparent horizon. The $|J| \le 2\,M_H^2$ condition is also equivalent to requiring a non-negative 
black hole surface gravity. We present numerical experiments of an accreting black 
hole that temporarily violates this extremality inequality. The initial 
configuration consists of a single, rotating black hole surrounded by a thick, 
shell cloud of negative energy density. For these numerical experiments, we introduce a new matter-without-matter 
evolution method.

\end{abstract}

\maketitle

\section{Introduction} 

Kerr spacetimes, representing a spinning \bh{} in isolation, have a bound on the maximum allowed angular momentum. 
If $M$ is the mass of the \bh{} and $J$ its angular momentum, the bound or \emph{extremality condition} reads $|J| \le M^2$. 
An extremal Kerr \bh{} saturates this condition (i.e. $|J| = M^2$). Kerr spacetimes violating the extremality condition (i.e. $|J|  > M^2$) have
naked singularities instead of \bh{s}. A natural question to ask then is whether more general \bh{} 
spacetimes (e.g. \bh{s} in the presence of matter, Maxwell fields, gravitational radiation and/or other \bh{s})  are also subject 
to extremality conditions, thus precluding the existence of \emph{super-extremal} \bh{s}. 

The existence of extremality conditions in more general \bh{} spacetimes has been investigated in a series 
of papers by Ansorg and collaborators. These studies considered axisymmetric and stationary \bh{s} 
surrounded by matter~\cite{2008CQGra..25c5009A,2008CQGra..25p2002H}. In some cases, Maxwell fields were also included~\cite{2010CMaPh.293..449H}.
Of direct relevance to our work is the study by Ansorg and Petroff ~\cite{2005PhRvD..72b4019A} demonstrating the feasibility of configurations with super-extremal \bh{s}. 
More recently, in a study of binary \bh{} initial data with nearly extremal spins~\cite{Lovelace:2008tw}, initial data with 
super-extremal \bh{s} were constructed; however, the \bbh{s} were enclosed 
by an \ahz{} with sub-extremal properties, thus effectively acting as a single, sub-extremal \bh{.}  

As pointed out by Booth and Fairhurst~\cite{Booth:2007wu}, at the core of any study of \bh{} extremality conditions is characterizing
what one means by mass and angular momentum associated with the \bh{,} namely distinguishing the ``local'' properties of the \bh{} from those of its environment. 
As shown also in~\cite{Booth:2007wu}, the natural tools for such a task are those provided by the isolated and dynamical horizon framework~\cite{2004LRR.....7...10A}. 
Using this framework, Booth and Fairhurst~\cite{Booth:2007wu} reformulated extremality conditions in terms of restrictions on isolated and dynamical horizons, and
introduced a parameter that determines how close a horizon is to extremality. 

The goal of our work is to investigate the formation of a \emph{super-extremal} \bh{} in a dynamical setup using the tools of numerical relativity. 
Specifically, we consider a sub-extremal \bh{} surrounded with a spherically symmetric cloud of negative energy density. As the \bh{} accretes the cloud, 
its mass decreases and its angular momentum increases. As a consequence, the dimensionless spin parameter of the \bh{} increases.
This approach to spinning up a \bh{} by decreasing its mass is complementary to the more astrophysically 
relevant case in which a \bh{}  spins up by gaining angular momentum from the material it swallows~\cite{1974ApJ...191..507T}. 
For clouds with enough negative energy density, we are able to build super-extremal \bh{s}. The super-extremal state is, however, not stable. 
Non-axisymmetric instabilities develop, triggering emission of gravitational radiation that carries with it copious amounts of angular momentum.  

Our study also introduces a new evolution scheme called \mwm{}.
The \mwm{} approach consists of evolving the spacetime geometry
using the BSSN~\cite{1999PhRvD..59b4007B} evolution equations without their matter source terms, namely the vacuum version of the evolution equations. 
The matter fields are constructed at each step from the BSSN constraints (Hamiltonian, momentum and connection constraints).
In addition, the stress energy tensor is required to have a form for which the matter source terms in the BSSN evolution equations vanish.
The approach is reminiscent of the constraint-violating \bh{} initial data evolved in a previous study~\cite{Bode:2009fq}. 
Our \mwm{} method differs, however, from work with hydro-without-hydro evolutions~\cite{1999PhRvD..60h7501B,2006CQGra..23S.579S}. In those studies, the matter hydrodynamics was
pre-prescribed and was used as input to construct the source terms in the evolution equations. 

The paper is organized as follows: In Section~\ref{sec:econd}, we summarize the extremality conditions and discuss the condition used in our 
numerical experiments to identify the emergence of super-extremality. 
Section~\ref{sec:mwm} describes the \mwm{} evolution approach and the conditions that the matter content must satisfy. 
In Sec.~\ref{sec:setup}, we describe the initial data configurations used in our simulations.
Results showing the formation of a super-extremal \bh{} are given in Sec.~\ref{sec:results}. In Sec.~\ref{sec:null}, 
we address the degree to which the null energy condition is satisfied by 
our spacetimes. In Sec.~\ref{sec:constraints} ,we investigate the late-time behavior in terms of the constraints. Conclusions  are presented in
Sec.~\ref{sec:end}. The numerical simulations and results were obtained with the Maya Code as described
in Refs.~\cite{2009arXiv0905.3914H,2009arXiv0912.0087B,Bode:2009fq}. Subscripts $a,b,c,..$ will denote spacetime indices while $i,j,k,...$ will denote spatial indices. We use units in which
$G=c=1$.

\section{Extremality Conditions}
\label{sec:econd}

There are three notions of extremality in stationary, asymptotically flat \bh{} 
spacetimes (see~\cite{Booth:2007wu} for details): 
i) the angular momentum $J$ and mass $M$ of a \bh{} satisfy the bound $\chi \equiv |J|/M^2 \le 1$;
ii) the surface gravity $\kappa$ of a \bh{} satisfies the bound $\kappa \ge 0$; and 
iii) the interior of a non-extremal \bh{} contains trapped surfaces, while 
there are no trapped surfaces in the interior of an extremal \bh{.}

The first notion of extremality ($\chi \le 1$) compares the Christodoulou 
mass~\cite{Christodoulou:1970wf} $M$ to the angular momentum $J$ of the \bh{.} The Christodoulou 
mass is given by
\begin{equation}
\label{eqn:massc}
M^2 \equiv M^2_H + \frac{J^2}{4\,M_H^2} = M^2_H\left[1 + \left(\frac{J}{2\,M^2_H}\right)^2\right]\,.
\end{equation} 
$M_H$ is a local measure of the \bh{} mass, called the irreducible mass. The mass is
computed from $M_H = \sqrt{A/16\pi}$, with $A$ the area of the \ahz{} of 
the \bh{.} 
The angular momentum $J$ is also obtained locally using approximate rotational
Killing vectors on the \ahz{}~\cite{2004LRR.....7...10A}.

For stationary \bh{s}, the surface gravity is given by~\cite{2004LRR.....7...10A}
\begin{equation}\label{eqn:surfgrav}
\kappa = \frac{1}{2\,M_H} \frac{(1-\chi^2)^{1/2}}{[1+(1-\chi^2)^{1/2}]}\,.
\end{equation}
Evident from Eq.~(\ref{eqn:surfgrav}) is the equivalence between the first two notions of extremality, namely
$\chi \le 1$ and $\kappa \ge 0$. 

The goal of our study is to investigate conditions under which the extremality conditions $\chi \le 1$ and $\kappa \ge 0$ are violated.
Using $\chi$ for this purpose has, however, the following drawback. 
From Eq.~(\ref{eqn:massc}), it is not difficult to show that
\begin{equation}
\left(\frac{2\,M_H^2}{M^2} - 1\right)^2 = 1-\frac{J^2}{M^4} = 1-\chi^2 \,.
\end{equation}
Thus, by construction $\chi \le 1$.
It has been then suggested~\cite{Booth:2007wu, Lovelace:2008tw} 
that a quantity more suitable to investigate extremality is $\zeta \equiv |J|/(2 M_H^2)$.
In terms of $\zeta$, the spin parameter $\chi$ and the surface gravity $\kappa$ take the form
\begin{eqnarray}
	\chi &=& 1 - \frac{(1-\zeta)^2}{1+\zeta^2}\\
	\kappa &=& \frac{1-\zeta^2}{4\,M}\,,
\end{eqnarray}
respectively. Thus, for $\zeta \le 1$, one recovers the extremality conditions $\chi \le 1$ and $\kappa \ge 0$. The advantage of the new spin parameter $\zeta$ is that
values $\zeta > 1$ are allowable. 
Furthermore, for $\zeta > 1$, one still has $\chi \le 1$, but 
the surface gravity, on the other hand, becomes negative. 

In Ref.~\cite{Booth:2007wu}, a new notion of extremality was introduced in terms of
$\zeta$:  A \bh{} is said to be \emph{sub-extremal} if $\zeta < 1$ ($\kappa > 0$),  \emph{extremal} if $\zeta = 1$ 
($\kappa = 0$), and \emph{super-extremal} if $\zeta > 1$ ($\kappa < 0$). 
This new notion of extremality is
derived assuming that:  i)  the spacetime is axisymmetric; 
ii) the null energy condition is satisfied; and iii) the cross sections of the horizon are embeddable in Euclidean $\mathbb{R}^3$.
A generalization of extremality, relaxing the axisymmetry and embeddability assumptions but keeping the null energy condition, is also possible~\cite{Booth:2007wu}.  

\section{Matter-without-Matter Evolutions} 
\label{sec:mwm}

Our study introduces also an evolution scheme that we call the \emph{matter-without-matter} method. 
Under the \mwm{} approach, the geometry of the spacetime is evolved  
using the vacuum or source-free BSSN evolution equations. There is no need for matter evolution equations. As we shall show next, 
the matter fields are obtained at every step of the evolution  from the 
BSSN constraints (Hamiltonian, momentum and connection constraints) together with \emph{equations-of-state}.
To demonstrate how the \mwm{} works, we will re-derive the BSSN evolution equations, taking
explicitly into account a matter content that is \emph{invisible} to these equations. 

The BSSN formulation of the Einstein equations
consists of a set of evolution equations for the conformal metric $\tilde \gamma_{ij}$, 
the conformal factor $\phi$, the trace of the physical extrinsic curvature $K$, the
trace-free part of the conformal extrinsic curvature $\tilde A_{ij}$, and the connection 
$\widetilde{\Gamma}^i \equiv \tilde{\gamma}^{jk} \widetilde{\Gamma}^i_{jk}=-\partial_j\tilde \gamma^{ij}$ 
(see Ref.~\cite{2010nure.book.....B} for details). The evolution equations for
$\phi$, $\tilde \gamma_{ij}$ and  $K$ are respectively:
\begin{eqnarray}
\label{eqn:phi}
\partial_o \phi &=& -\frac{1}{6}\,\alpha\,K\\
\label{eqn:hij}
\partial_o \tilde \gamma_{ij} &=& -2\,\alpha\,\tilde A_{ij} \\
\label{eqn:K}
\partial_o K &=& -\nabla_i\nabla^i \alpha + \alpha\left(\tilde A^{ij}\tilde A_{ij} 
	      + \frac{1}{3} K^2\right)\nonumber\\
	     &+& 4\,\pi\,\alpha(\rho+S)\,,
\end{eqnarray}
$\nabla$ denotes covariant differentiation with respect to the physical metric $\gamma_{ij}=e^{4\phi}\tilde\gamma_{ij}$,
$\alpha$ is the lapse function, and $\beta^i$ is the shift vector; we define
$\partial_o \equiv\partial_t - {\cal L}_\beta$, with ${\cal L}_\beta$ the spatial Lie derivative along $\beta^i$. Above, the source terms $\rho$ and $S = \gamma^{ij}S_{ij}$  
are obtained from 
\begin{eqnarray}
\rho &=& n^a\,n^b T_{ab}\\
j^i &=& -\gamma^{ia} n^b T_{ab} \\
S_{ij} &=& \gamma_i^a\gamma_j^b T_{ab}
\end{eqnarray}
with $T_{ab}$ the stress-energy tensor and with $n^a$ the unit, time-like normal to the constant $t$ hypersurfaces. That is, the stress-energy tensor has the following form:
\begin{equation}
\label{eqn:tmunu1}
T_{ab} = \rho\,n_a\,n_b + 2\, j_{(a}\,n_{b)} + S_{ab}\,.
\end{equation}

Before considering the evolution equations for $\tilde A_{ij}$ and
$\widetilde{\Gamma}^i$, we need to recall the constraints in the BSSN formulation. 
As with any 3+1 formulation of the Einstein field equations of general relativity, the BSSN formulation 
involves the Hamiltonian and momentum constraints. In terms of the BSSN variables,  these constraints read:
\begin{equation}
\label{eqn:hc}
e^{-5\phi}\widetilde\nabla_i\widetilde\nabla^i e^\phi -\frac{e^{-4\phi}}{8}\widetilde R + \frac{1}{8}\tilde A^{ij}\tilde A^{ij}- \frac{1}{12}K^2 
= -2\,\pi\,\rho
\end{equation}
and
\begin{equation}
\label{eqn:mc}
e^{-6\phi} \widetilde\nabla_j (e^{6\phi}\tilde A^{ij}) -\frac{2}{3}\widetilde\nabla^i K = 8\,\pi j^i\,,
\end{equation}
respectively. In addition, the BSSN formulation introduces the following new constraints:
\begin{eqnarray}
\label{eqn:ac}
{\cal A} &=& \tilde A^i\,_i = 0\\
\label{eqn:sc}
{\cal S} &=& \tilde\gamma - 1 = 0\\
\label{eqn:gc}
{\cal G}^i &=& \widetilde\Gamma^i + \partial_j\tilde\gamma^{ij} = 0\,.
\end{eqnarray}
Our Maya Code actively imposes the trace-free (\ref{eqn:ac}) and unit-determinant (\ref{eqn:sc}) constraints. On the other hand,
in numerical evolutions, the Hamiltonian (\ref{eqn:hc}), momentum (\ref{eqn:mc}) and connection (\ref{eqn:gc}) constraints are not 
explicitly imposed. One of the main virtues of the BSSN formulation is precisely that aspect. Namely, in the course of a BSSN evolution, the constraints  (\ref{eqn:hc}), (\ref{eqn:mc})
and  (\ref{eqn:gc}) are preserved within tolerable levels. In our \mwm{} scheme, the Hamiltonian, momentum and connection constraints 
play a different role. The Hamiltonian and momentum constraints are used to construct $\rho$ and $j^i$, respectively.
But more importantly, in \mwm{} simulations (see below), ${\cal G}^i$ evolves away from ${\cal G}^i = 0$ and provides a key ingredient in determining the dynamics 
of the matter content. For this reason, we need next to re-derive  the evolution equations for $\tilde A_{ij}$ and
$\widetilde{\Gamma}^i$ without imposing the ${\cal G}^i = 0$ constraint. 

For the $\tilde A_{ij}$ evolution equation, the only place where ${\cal G}^i$ enters is in the Ricci tensor $R_{ij}$, which is normally computed from 
from $R_{ij} =  \widetilde R_{ij} + R^\phi_{ij}$ with
\begin{eqnarray}
\label{eqn:tilderij}
   \widetilde R_{ij} &=& - \frac12 \tilde{\gamma}^{lm} \partial_l \partial_m \tilde{\gamma}_{ij} \nonumber\\
   	&+& \tilde{\gamma}^{lm} \left( 2 \widetilde{\Gamma}^k_{l(i} \widetilde{\Gamma}_{j)km}   + \widetilde{\Gamma}^k_{im} \widetilde{\Gamma}_{klj} \right)\nonumber\\
	&+& \tilde{\gamma}_{k(i} \partial_{j)} \widetilde\Gamma^k  + \widetilde\Gamma^k \widetilde{\Gamma}_{(ij)k} \,,
\end{eqnarray}
and
\begin{eqnarray}
\label{eqn:phirij}
R^\phi_{ij} &=& -2\left(\widetilde\nabla_i\widetilde\nabla_j\phi  + \tilde\gamma_{ij}\widetilde\nabla_k\widetilde\nabla^k\phi \right)\nonumber\\
&+& 4\left( \widetilde\nabla_i\phi\widetilde\nabla_j\phi - \tilde\gamma_{ij}\widetilde\nabla_k\phi\widetilde\nabla^k\phi  \right)\,,
\end{eqnarray}
respectively. If ${\cal G}^i \ne 0$, the  Ricci tensor $\widetilde R_{ij}$ acquires an additional term, namely
\begin{eqnarray}
\label{eqn:tilderij2}
   \widetilde R_{ij} &\rightarrow & \widetilde R_{ij} = - \frac12 \tilde{\gamma}^{lm} \partial_l \partial_m \tilde{\gamma}_{ij} \nonumber\\
   	&+& \tilde{\gamma}^{lm} \left( 2 \widetilde{\Gamma}^k_{l(i} \widetilde{\Gamma}_{j)km} 	  + \widetilde{\Gamma}^k_{im} \widetilde{\Gamma}_{klj} \right)\,\nonumber\\
	&+& \tilde{\gamma}_{k(i} \partial_{j)} \widetilde\Gamma^k  + \widetilde\Gamma^k \widetilde{\Gamma}_{(ij)k} \nonumber\\
	&-& \tilde{\gamma}_{k(i} \partial_{j)} {\cal G}^k  - {\cal G}^k \widetilde{\Gamma}_{(ij)k} \,.
\end{eqnarray}
Therefore, $R_{ij}$ is given instead as 
 $R_{ij} = \widetilde R_{ij} + R^\phi_{ij} + 8\,\pi \,M_{ij}$ with $\widetilde R_{ij}$ and $R^\phi_{ij}$ given by (\ref{eqn:tilderij}) and  (\ref{eqn:phirij}), respectively, and
\begin{equation}	
	M_{ij} \equiv -\frac{1}{8\,\pi}\left( \tilde{\gamma}_{k(i} \partial_{j)} {\cal G}^k  + {\cal G}^k \widetilde{\Gamma}_{(ij)k}\right)\,.
\end{equation}
With $M_{ij}$ viewed as an additional source term, one can follow the standard BSSN derivation~\cite{2010nure.book.....B} of the 
evolution equation for $\tilde A_{ij}$ and arrive at:
\begin{eqnarray}
\label{eqn:Aij2}
\partial_o \tilde A_{ij} &=& e^{-4\,\phi}\left[ -\nabla_i\nabla_j\alpha + \alpha\,R_{ij} - 8\,\pi\,\alpha(S_{ij}-M_{ij})\right]^{TF}\nonumber\\
&+& \alpha(K\,\tilde A_{ij} - 2\,\tilde A_{ik}\tilde A^{k}\,_{j})\,,
 \end{eqnarray}
where $TF$ denotes the trace-free part of the tensor.

The remaining equation to consider is the evolution equation for $\widetilde \Gamma^i $. 
Once again, we will derive this equation without the assumption that the constraint 
${\cal G}^i = 0$ holds. The starting point is the time derivative of $ \widetilde\Gamma^i =  - \partial_j\tilde\gamma^{ij}  + {\cal G}^i$:
\begin{equation}
\label{eqn:gdot}
\partial_t\widetilde\Gamma^i =  - \partial_j\partial_t\tilde\gamma^{ij}  + \partial_t{\cal G}^i\,.
\end{equation}
On the other hand, from Eq.~(\ref{eqn:hij}):
\begin{eqnarray}
\label{eqn:gij2}
\partial_t \tilde\gamma^{ij} &=& 2\alpha\,\tilde A^{ij}  +\beta^k\partial_k\tilde\gamma^{ij} \nonumber\\ 
&-& \tilde\gamma^{ik}\partial_k\beta^j - \tilde\gamma^{jk}\partial_k\beta^i + \frac{2}{3} \tilde\gamma^{ij}\partial_k\beta^k\,,
\end{eqnarray}
which after differentiation reads
\begin{eqnarray}
\label{eqn:dgij}
\partial_t \partial_j\tilde\gamma^{ij} &=& \partial_j(2\alpha\,\tilde A^{ij})  \nonumber\\
&+&\beta^k\partial_k\partial_j\tilde\gamma^{ij} - \partial_j\tilde\gamma^{jk}\partial_k\beta^i + \frac{2}{3}\partial_j \tilde\gamma^{ij}\partial_k\beta^k\nonumber\\ 
&-& \tilde\gamma^{jk}\partial_j\partial_k\beta^i - \frac{1}{3} \tilde\gamma^{ij}\partial_j\partial_k\beta^k \,.
\end{eqnarray}
Thus, Eq.~(\ref{eqn:gdot}) becomes
\begin{eqnarray}
\label{eqn:gdot2}
\partial_t\widetilde\Gamma^i &=& -2\,\tilde A^{ij}\partial_j\alpha -2\,\alpha\,\partial_j\tilde A^{ij}  \nonumber\\
&-&\beta^k\partial_k\partial_j\tilde\gamma^{ij} + \partial_j\tilde\gamma^{jk}\partial_k\beta^i - \frac{2}{3}\partial_j \tilde\gamma^{ij}\partial_k\beta^k\nonumber\\ 
&+& \tilde\gamma^{jk}\partial_j\partial_k\beta^i + \frac{1}{3} \tilde\gamma^{ij}\partial_j\partial_k\beta^k  + \partial_t{\cal G}^i\,.
\end{eqnarray}
Next we use the momentum constraint (\ref{eqn:mc}) to eliminate the term involving $\partial_j\tilde A^{ij}$, and rewrite (\ref{eqn:gdot2}) as:
\begin{eqnarray}
\label{eqn:gdot3}
\partial_t \widetilde \Gamma^i &=& -2\,\tilde A^{ij}\partial_j \alpha - 16\,\pi\alpha\tilde\gamma^{ij}j_j  \nonumber \\
&-& 2\,\alpha\left (-\widetilde\Gamma^i_{jk}\tilde A^{jk} - 6 \tilde A^{ij}\partial_j\phi +\frac{2}{3}\tilde\gamma^{ij}\partial_j K\right) \nonumber\\
&-&\beta^k\partial_k\partial_j\tilde\gamma^{ij} + \partial_j\tilde\gamma^{jk}\partial_k\beta^i - \frac{2}{3}\partial_j \tilde\gamma^{ij}\partial_k\beta^k\nonumber\\ 
& +& \tilde\gamma^{jk}\partial_j\partial_k\beta^i + \frac{1}{3}  \tilde\gamma^{ij}\partial_j\partial_k\beta^k + \partial_t{\cal G}^i\,.
\end{eqnarray}
With the help of $\partial_j\tilde\gamma^{ij} = -\widetilde\Gamma^i + {\cal G}^i$, we obtain
\begin{eqnarray}
\label{eqn:gdot4}
\partial_t \widetilde \Gamma^i &=& -2\,\tilde A^{ij}\partial_j \alpha - 16\,\pi\alpha\tilde\gamma^{ij}j_j  \nonumber \\
&-& 2\,\alpha\left (-\widetilde\Gamma^i_{jk}\tilde A^{jk} - 6 \tilde A^{ij}\partial_j\phi +\frac{2}{3}\tilde\gamma^{ij}\partial_j K\right) \nonumber\\
&+& \beta^j\partial_j\widetilde\Gamma^i - \widetilde\Gamma^j\partial_j\beta^i + \frac{2}{3}\widetilde\Gamma^i\partial_j\beta^j\nonumber \\
&-& \beta^j\partial_j{\cal G}^i + {\cal G}^j\partial_j\beta^i - \frac{2}{3}{\cal G}^i\partial_j\beta^j\nonumber \\
& +& \tilde\gamma^{jk}\partial_j\partial_k\beta^i + \frac{1}{3}  \tilde\gamma^{ij}\partial_j\partial_k\beta^k + \partial_t{\cal G}^i\,,
\end{eqnarray}
which can be rewritten as
\begin{eqnarray}
\label{eqn:gamma}
\partial_o \widetilde \Gamma^i &=& -2\,\tilde A^{ij}\partial_j \alpha  \nonumber \\
&-& 2\,\alpha\left (-\widetilde\Gamma^i_{jk}\tilde A^{jk} - 6 \tilde A^{ij}\partial_j\phi +\frac{2}{3}\tilde\gamma^{ij}\partial_j K\right) \nonumber\\
& +& \tilde\gamma^{jk}\partial_j\partial_k\beta^i + \frac{1}{3}  \tilde\gamma^{ij}\partial_j\partial_k\beta^k \nonumber \\
& +& \partial_o{\cal G}^i - 16\,\pi\alpha\tilde\gamma^{ij}j_j \,,
\end{eqnarray}
with both $\widetilde \Gamma^i$ and ${\cal G}^i$ treated as vector densities of weight 2/3.
Equations (\ref{eqn:phi}), (\ref{eqn:hij}), (\ref{eqn:K}), (\ref{eqn:hc}), (\ref{eqn:mc}), (\ref{eqn:Aij2}) and (\ref{eqn:gamma}) constitute the 
basis of our \mwm{} method. 

From the form of the stress-energy tensor (\ref{eqn:tmunu1}), we see that without any further assumptions, our set of equations is 
not able to determine $S_{ij}$. The essence of a \mwm{} evolution is then to impose that matter evolves in such a way that the source terms in
(\ref{eqn:K}), (\ref{eqn:Aij2}) and (\ref{eqn:gamma}) vanish, namely 
\begin{eqnarray}
\label{eq:Scond}
\rho + S &=& 0\\
\label{eq:Sijcond}
(S_{ij} - M_{ij})^{TF} &=& 0\\
 \partial_o{\cal G}^i - 16\,\pi\alpha e^{-4\,\phi}j^i &=&0\,.
\end{eqnarray}
These conditions imply that the stress-energy tensor (\ref{eqn:tmunu1})  takes the following form:
\begin{equation}
\label{eqn:tmunu}
T_{ab} = \rho\,\left(n_a\,n_b -\frac{1}{3}\gamma_{ab}\right)+ 2\, j_{(a}\,n_{b)} + \gamma_a^i\gamma_b^{j}M^{TF}_{ij}\,.
\end{equation}

The obvious advantage of the \mwm{} method is the direct use of a vacuum (e.g. black hole) BSSN evolution code to construct the spacetime geometry.
The matter fields $\rho$, $j^i$ and ${\cal G}^i $ are obtained after every step from the Hamiltonian (\ref{eqn:hc}), momentum (\ref{eqn:mc}) and connection (\ref{eqn:gc}) constraints.  
Since the \mwm{} method can be also viewed as evolving constraint violating data, there are no guarantees that the method is capable of producing stable evolutions.
A general proof of the conditions that the initial data have to satisfy in order for the \mwm{} to yield stable evolutions is beyond the scope of this study. 
We have found, however, that the set of configurations and evolutions for the present study on super-extremality were all numerically stable.

\section{Initial Setup}
\label{sec:setup}

Our study consists of a single, 
spinning \bh{} puncture~\cite{1997PhRvL..78.3606B}  enclosed by
a thick, spherically symmetric, shell cloud of (in most cases negative) energy density.  
The shell surrounding the \bh{} has a Gaussian profile and is initially
static, with stress-energy tensor given by $T_{ab} = \rho\,(n_a\,n_b -\gamma_{ab}/3)$. Specifically, with
this choice of stress-energy tensor, initially $j^i = {\cal  G}^i  = \partial_o{\cal  G}^i = 0$.
 
Vanishing $j^i$ implies that the momentum constraint reduces to the vacuum case.
One can thus directly use the Bowen and York extrinsic curvature solutions for a spinning 
puncture~\cite{1980PhRvD..21.2047B}:
\begin{equation}
\label{eqn:by}
\hat A_{ij} = -\frac{3}{r^3}\left(\epsilon_{ilk}\,\hat r_j + \epsilon_{jlk}\,\hat r_i\right) \hat r^i J^k\,,
\end{equation} 
with $\hat r^i$ the radial unit vector and $J^i$ the puncture's angular momentum. 
Constructing initial data requires then solving only the Hamiltonian constraint~\cite{1979sgrr.work...83Y}
\begin{equation}
\label{eqn:hcpsi}
 \Delta\psi +  \frac{1}{8}\psi^{-7} \tilde A_{ij} \tilde A^{ij}  + 2\,\pi\psi^5\rho = 0
\end{equation}
with $\hat A_{ij}$ given by (\ref{eqn:by}) and $\Delta$ the flat Laplacian. Notice that the standard assumptions of conformal flatness and vanishing trace of the extrinsic curvature have been used.
Once a solution to Eq.~(\ref{eqn:hcpsi}) is found, the initial data for the spatial metric and the extrinsic curvature are obtained from $\gamma_{ij} = \psi^4\,\eta_{ij}$ and $K_{ij} = \psi^{-2}\tilde A_{ij}$, respectively.

To solve Eq.~(\ref{eqn:hcpsi}), we use the puncture ansatz $\psi = 1+ m_p/(2\,r)+u$, with
$m_p$ the puncture's bare mass parameter. 
We choose the source $\rho$ as:
\begin{equation}
\label{eqn:rhosource}
\rho = \rho_o e^{-(r-r_o)^2/\sigma^2} \left(1+\frac{m_p}{2\,r}\right)^{-5}\,,
\end{equation}
and the bare angular momentum as $J^i = J\,\hat z^i$.  For all our simulations, we set $J=0.8$.
The last factor in (\ref{eqn:rhosource}) is needed for regularity at the location of the puncture.
Thus,  Eq.~(\ref{eqn:hcpsi})
becomes:
\begin{eqnarray}
\label{eq:hc2}
 &&\Delta u +  \frac{18}{8}\frac{J^2\,r\,\sin^2\theta}{(r+m_p/2+u\,r)^7 } \nonumber\\
 &&+ 2\,\pi\left(\frac{r+m_p/2+u\,r}{r+m_p/2}\right)^5\rho_o e^{-(r-r_o)^2/\sigma^2}  = 0\,,
\end{eqnarray}
with $\hat r^i\,\hat z^j \,\eta_{ij} = \cos{\theta}$.

The parameters $r_o$ and $\sigma$ are chosen to favor the accretion of most of the shell by the \bh{} 
in evolution time-scales of $\lesssim 40\,M$. 
The energy density amplitude parameter $\rho_o$ is the \emph{knob} that controls the mass associated 
with the shell, and thus regulates the total ADM
mass $M_o$ in the initial data. 

Our approach to breaking the extremality bound $\zeta \equiv J/(2\,M_H^2) = 1$ is to not only 
increase the angular momentum $J$ of the \bh{,} but most importantly to decrease 
its mass $M_H$. To accomplish this, we endow the shell surrounding the \bh{} with a negative mass, which 
decreases the \bh{'s} mass as it gets accreted.

It should be noted that in isolation any spinning puncture that models a rotating \bh{} has a small ($\lesssim 0.1\%$ 
of the total energy) spurious amount of gravitational radiation. This spurious radiation carries away a burst of angular 
momentum~\cite{Gleiser:1997ng}; this amount, however, is small and does not affect the conclusions from our numerical experiment.

\begingroup
\begin{center}
\begin{table}[ht]
\begin{ruledtabular}
\begin{tabular}{c|ccccc}
  Run & $r_o/M_o$ & $\sigma/M_o$ & $\rho_o\,M_o^2$ & $m_p/M_o$ & $\chi_o$  \\
\hline
  V0  &  --  &  --  & 0                    & 0.62 & 0.80 \\ 
  V1  & 0.88 & 0.88 & $+7.03\times10^{-3}$ & 0.55 & 0.62 \\ 
  V2  & 1.03 & 1.03 & $-3.77\times10^{-4}$ & 0.64 & 0.85 \\ 
  V3  & 1.07 & 1.07 & $-1.05\times10^{-3}$ & 0.71 & 0.90 \\ 
  V4  & 1.18 & 1.18 & $-1.72\times10^{-3}$ & 0.72 & 1.12 \\ 
  V5  & 1.23 & 1.23 & $-2.00\times10^{-3}$ & 0.76 & 1.20 
\end{tabular}
\end{ruledtabular}
\caption{Initial parameters defining the matter shell ($r_o, \sigma, \rho_o$). The last two columns display the puncture mass parameter $m_p$ and 
ADM spin parameter $\chi_o = J_o/M_o^2$. All parameters are given in units of the ADM mass $M_o$ }
\label{tab:cv_id}
\end{table}
\end{center}
\endgroup

Table \ref{tab:cv_id} shows the parameters used for our simulations: the shell parameters ($r_o, \sigma, \rho_o$) and the puncture mass $m_p$.
These quantities and further results are given 
in dimensionless units in terms of the ADM mass $M_o$. 
The last column in Table \ref{tab:cv_id} reports for each simulation the total spin parameter $\chi_o\equiv J_o/M_o^2$, with $J_o$ the total ADM angular momentum.
Since $\chi_o$ is associated with the ADM mass and angular momentum of the spacetime, it is not subject to an extremality condition.
In particular, $\chi_o$ includes angular momentum from outside the horizon.
Notice that case V0 consists of a single puncture in vacuum. This case is used as a control run. Case V1 is 
a fiducial evolution with a {\it positive} energy density shell. Cases V2--V5 contain shells with increasingly negative energy density. 
They are the central piece of our study.
Table~\ref{tab:cv_id2} gives the initial values for the irreducible $M_H$ and Christodoulou $M$ masses for each case. 
The table also shows the initial values of the \bh{'s} spin parameters $\chi= J/M^2$, $\zeta= J/(2\,M_H^2)$, and its surface gravity $\kappa$.

\begingroup
\begin{center}
\begin{table}[ht]
\begin{ruledtabular}
\begin{tabular}{c|ccccc}
  Run & $M_H/M_o$ & $M/M_o$ & $\chi$ & $\zeta$ & $\kappa\,M_o$ \\
\hline
  V0  & 0.886 & 0.994 &0.810 & 0.511 & 0.209 \\ 
  V1  & 0.794 & 0.886 &0.796 & 0.496 & 0.237 \\ 
  V2  & 0.908 & 1.021 &0.814 & 0.515 & 0.202 \\ 
  V3  & 0.958 & 1.071 &0.800 & 0.499 & 0.196 \\ 
  V4  & 1.020 & 1.156 &0.830 & 0.533 & 0.175 \\ 
  V5  & 1.056 & 1.120 &0.835 & 0.538 & 0.168 
\end{tabular}
\end{ruledtabular}
\caption{Initial irreducible mass $M_H$, Christodoulou mass $M$, dimensionless spin parameters $\chi = J/M^2$ and $\zeta = J/(2\,M_H^2)$,
and surface gravity $\kappa$ of the \bh{} in units of the total ADM mass $M_o$}
\label{tab:cv_id2}
\end{table}
\end{center}
\endgroup

Although we are dealing with a single \bh{,} the numerical simulations are challenging. 
As the \bh{} approaches extremality, the horizon undergoes extreme pancake-like deformation. Because of this severe deformation, in order to capture the dynamics 
in the vicinity of the \bh{} and to have a chance of locating its \ahz{,} we were forced to use meshes with $104^2\times52$ 
grid point shapes, and resolutions in the finest mesh of at least  $M/200$, in addition to using 6th-order accurate finite differencing.
For some of the cases, we carried out simulations with resolutions of  $M/167$ in the finest mesh to investigate the dependence of our results with resolution. 
We did not find noticeable differences regarding the onset of super-extremality. Resolution effects were mostly visible in tracking the \ahz{.}  

\section{Super-extremal Black Holes}
\label{sec:results}

\begin{figure}[ht]
\includegraphics[width=0.45\textwidth]{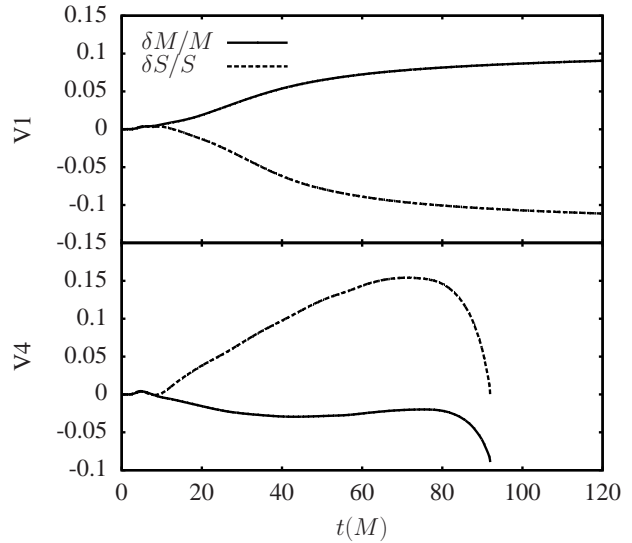}
\caption{ Fractional change of the Christodoulou mass (solid) and spin (dotted) for the cases of V1 (top) and V4 (bottom).}
\label{fig:MassSpin}
\end{figure}

We evolved the series of runs described in Table~\ref{tab:cv_id}. 
As previously stated, the case V0 is a control simulation to show that in the absence of matter, the \bh{} mass and angular momentum indeed remain constant. 
For all the non-vacuum cases, the shell is absorbed early in the evolution. As a consequence,
the mass $M_H$ of the \bh{} changes, increasing for the positive energy case V1 and decreasing for the remaining negative energy cases V2-V5. 

The case V1 involving a positive energy density shell demonstrates an important feature of our data: 
the shell of matter has negative angular momentum and positive mass (see Figure \ref{fig:MassSpin}). 
The \bh{} mass increases and the spin decreases as the shell is absorbed.
(The time axis in all figures is given in units of the initial Christodoulou 
mass of the \bh{.}) Figure~\ref{fig:MassSpin} also shows the behavior of the \bh{} mass and angular 
momentum for the V4 case with a negative energy shell. 
The Equivalence Principle indicates that motion of test bodies 
is independent of their mass (even the sign of their mass). While our
shells are not test bodies, we do expect the motion of the shells for the 
V1 positive mass case and the V4 negative mass case to be similar. 
We thus anticipate and find, for example in V4, that as the 
negative mass shell falls into the \bh{,} the \bh{} mass {\it decreases} and the \bh{} angular 
momentum {\it increases} (see Figure \ref{fig:MassSpin}), thus indicating that 
our negative energy shells have initially \emph{positive} angular momentum. 

\begin{figure}[ht]
\includegraphics[width=0.45\textwidth]{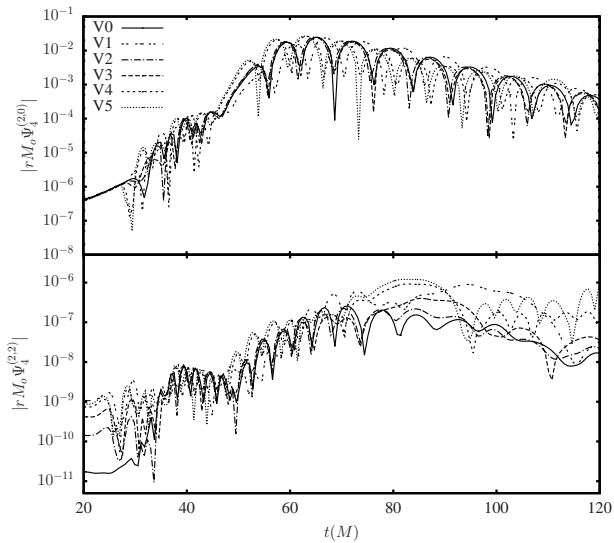}
\caption{Modes (2,0) and (2,2) of the Weyl scalar $\Psi_4$ for all cases in the first $120\,M$ of evolution time.}
\label{fig:gws}
\end{figure}

\begin{figure}[ht]
\includegraphics[width=0.45\textwidth]{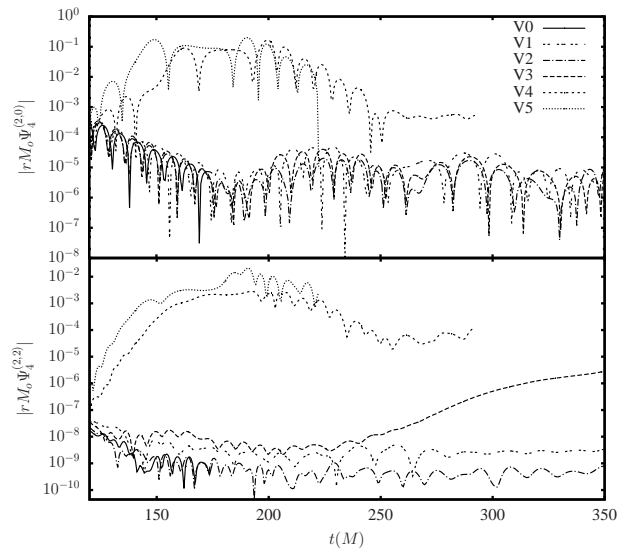}
\caption{Modes (2,0) and (2,2) of the Weyl scalar $\Psi_4$ for all cases after $150\,M$ of evolution time.}
\label{fig:gwslate}
\end{figure}

Although the initial configuration is axisymmetric, there is in all cases gravitational radiation emitted during the evolution. 
The first emission occurs early on around $t \sim 40\,M$. This is due to the spurious gravitational radiation in spinning punctures mentioned in the previous section.
Details of this emission can be seen in Figure~\ref{fig:gws} where we 
show the (2,0) and (2,2) modes of the Weyl scalar $\Psi_4$.
In addition, the accretion induces non-axisymmetric deformations that trigger a larger burst around $t \sim 60\,M$, followed by quasi-normal ringing (see top panel Fig.~\ref{fig:gws}).
Notice from the bottom panel of Fig.~\ref{fig:gws} that the (2,2) mode, which could potentially carry angular momentum, is slightly stronger for the V4 and V5 cases.
At late times, $t > 150\,M$, non-axisymmetric instabilities trigger a much stronger additional burst of radiation for those two cases, as seen in Figure~\ref{fig:gwslate}.
As a consequence, the \bh{} looses angular momentum.

\begin{figure}[ht]
\includegraphics[width=0.45\textwidth]{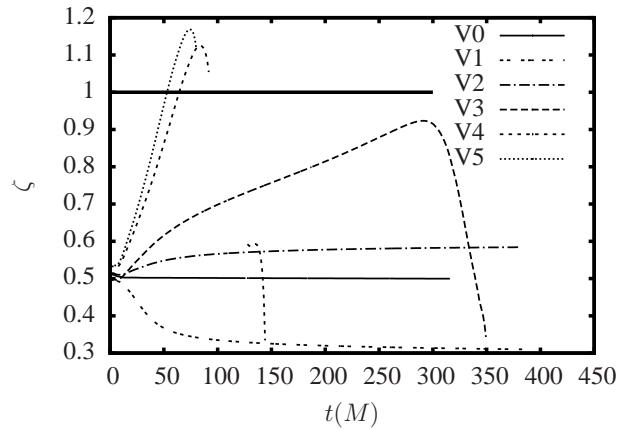}
\caption{Spin parameter $\zeta$ as a function of evolution time.}
\label{fig:Zeta}
\end{figure}

Figure~\ref{fig:Zeta} shows the core of our study, namely the evolution of the spin parameter $\zeta$ during the spacetime evolution.
As expected, $\zeta$ is constant in the V0 case. Also expected in case V1, with a positive energy density shell, is the decrease of the spin parameter $\zeta$ of the \bh{}
due to the monotonic increase (decrease) of its mass (spin) observed in Fig.~\ref{fig:MassSpin}.

Cases V2--V5, with their negative energy density shells, are our main focus.
As the shell is swallowed, the mass of the \bh{} decreases monotonically since it accretes negative mass. Furthermore, given that our negative energy shells
have positive angular momentum, the spin of the \bh{} will monotonically increase during the accretion.  When taken together, the increase in spin and decrease in mass
yield the observed growth of $\zeta$ in Fig.~\ref{fig:Zeta}.
Notice from Fig.~\ref{fig:Zeta} that the larger the negative energy density of the shell, the faster the increase experienced by $\zeta$.
Also, only the cases V4 and V5 breach the extremality bound $\zeta = 1$. These cases are the ones in which $\chi_o > 1$ initially (see Table \ref{tab:cv_id}).
V4 becomes super-extremal at a time $\sim 65\,M$ and V5 at $\sim 45\,M$.
In both cases, $\zeta$ continues to grow until it reaches a maximum, at $\sim 85\,M$ for V4 and $\sim 70\,M$ for V5.
The drop in $\zeta$ after reaching the maximum in V4 and V5 is due to the emission of angular momentum. The \bh{} is not able to sustain
super-extremality and simultaneously retain axisymmetry.

Fig.~\ref{fig:Zeta} also shows that after V4 and V5 reach a maximum, the $\zeta$ lines terminate.  This signals the time when we are no longer able to locate the \ahz{} of the \bh{.}
We have investigated whether the loss of the \ahz{} is real or due to numerical effects.
Simulations at different resolutions tells us that most likely the loss of the horizon is a numerical artifact:  the last horizon times depend on resolution.
The horizon undergoes a severe pancake deformation that the \ahz{} tracker is not able to handle because of the lack of resolution.
We stress that the simulation does not crash; we only lose the horizon. This should not be viewed as ``strange'' since it is well known that stable puncture
simulations do not resolve the puncture nor its immediate vicinity. If the horizon is too small, as with rapidly spinning \bh{s}, the horizon is in danger of entering the under-resolved
region near the puncture.

Notice in Fig.~\ref{fig:Zeta} that for V4 around $150\,M$, when $\zeta$ becomes again subcritical, we are again able to find the \ahz{,} albeit only briefly.
Although the value of  $\zeta$ has significantly decreased, the shape of the horizon remains pancake-like, thus the challenge of locating the horizon remains.
Since the \ahz{} shape is coordinate dependent, we are currently investigating modifications to the puncture gauges
that alleviate the horizon deformations, giving us a better chance of locating the horizon.

Finally, it is evident in Fig.~\ref{fig:Zeta} the increase that $\zeta$ experiences in the sub-extremal cases V2 and V3. V2 gives hints of reaching an asymptotic value at later times.
On the other hand, V3 has an approximate linear growth between $75\,M$ and $300\,M$ when it saturates at $\zeta \sim 0.92$.

\section{Null Energy Condition}
\label{sec:null}

\begin{figure}[ht]
\includegraphics[width=0.45\textwidth]{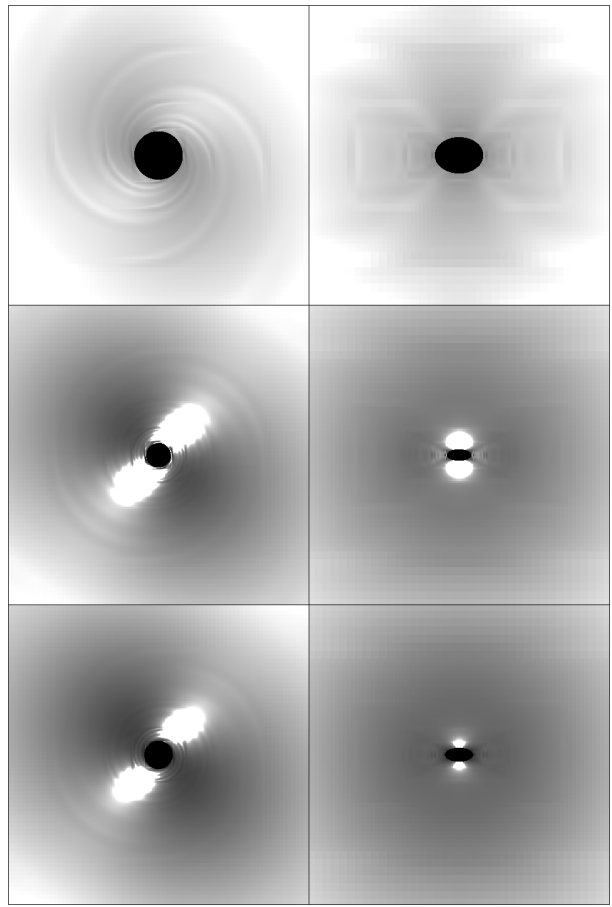}
\caption{Snapshots of ${\cal N}$ computed from Eq.~(\ref{eq:nec3}). From top to bottom,  case V3, V4 and V5 are shown. 
Left panels show a cut through the equatorial plane, and right panel along the rotation axis of the \bh{.} White areas are regions with positive values of ${\cal N}$, and gray shaded areas those in which the
null energy condition is violated. Snapshots are at times of $65.28M$, $62.72M$, and $50.24M$ for V3, V4 and V5 
respectively and cover a region $6M$ across. The grayscale is logarithmic in absolute value with a global minimum of -0.005. }
\label{fig:null}
\end{figure}

\begin{figure}[ht]
\includegraphics[width=0.45\textwidth]{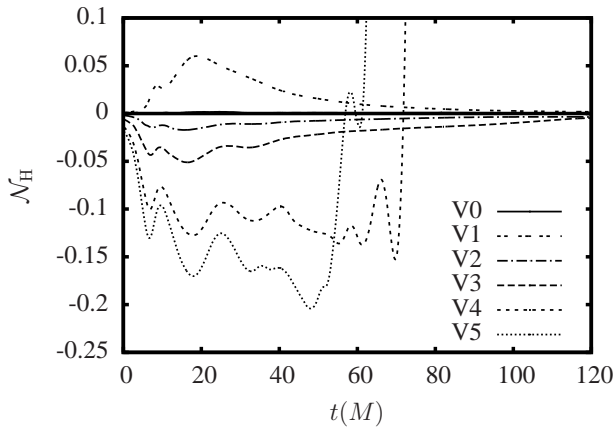}
\caption{Average of the null energy condition ${\cal N}_H$ on the surface of the \ahz{.}}
\label{fig:NullEC}
\end{figure}

As mentioned in Sec.~\ref{sec:econd}, one of the assumptions that the extremality condition $\zeta \le 1$ hinges on is the validity of the null energy condition.
The null-energy condition states that for all null vectors $k^a$
\begin{equation}
\label{eqn:nec}
{\cal N} = T_{ab}k^ak^b \ge 0\,.
\end{equation}
Substitution in Eq.~(\ref{eqn:nec}) of the stress-energy tensor $T_{ab}$ with the form given by Eq.~(\ref{eqn:tmunu})  yields  
\begin{eqnarray}
\label{eq:nec2}
{\cal N} &=& \frac23\,\rho - 2\,j_a\,l^a + M_{ab}^{TF}l^a\,l^b\,.
\end{eqnarray}
Above, we have used a null vector $k^a = n^a + l^a$ with $l^a n_a = 0$ and $l^al_a = 1$.
If we choose $l^a$ to be radial and centered at the \bh{,} i.e. $l^a = \hat e^a_r$, then
\begin{eqnarray}
\label{eq:nec3}
{\cal N} &=& \frac23\,\rho -2\,j_r + M_{rr}^{TF}\,,
\end{eqnarray}
We have evaluated (\ref{eq:nec3}) throughout the computational domain. Not surprisingly, we found that for all the negative energy density cases the null energy condition is
violated in regions near the \bh{.} Figure~\ref{fig:null} shows for cases V3, V4 and V5 snapshots
of both the equatorial plane (left panels) and the rotation axis (right panels) of the \bh{.}
White areas are regions with positive values of ${\cal N}$, and gray shaded areas those in which the
null energy condition is violated. The grayscale is logarithmic in absolute value, with a global minimum of -0.005.

Notice in the left panels in Fig.~\ref{fig:null} the evident inspiral structure in ${\cal N}$ associated with the accretion of the
negative energy shell. Also interesting is that among the three cases depicted, only the one with the weakest negative energy shell
shows \emph{no} regions where the null energy condition is \emph{not} violated.

We have also evaluated the null energy condition at the surface of the \ahz{.} That is, we evaluated ${\cal N}_H=T_{ab} k^a\,k^b$ on 
the \ahz{} using the null vector $k^a  = n^a + s^a$, 
with $s^a$ the 
spatial unit normal to the \ahz{.}  
In Fig.~\ref{fig:NullEC}, we show how the average of ${\cal N}_H$ changes during the course of the evolutions.
The null energy condition is clearly violated in cases V2--V5. A closer look at the data shows that this null 
energy condition is violated not only on average, but also everywhere on the \ahz{} surface.

In summary, the violations of the null energy condition in the vicinity of the \bh{} (Fig.~\ref{fig:null}) and on the surface of the \ahz{} (Fig.~\ref{fig:NullEC})
demonstrates consistency with the violation of the extremality bound $\zeta = 1$.

\section{Constraints and Late Behavior}
\label{sec:constraints}

\begin{figure}[ht]
\includegraphics[width=0.45\textwidth]{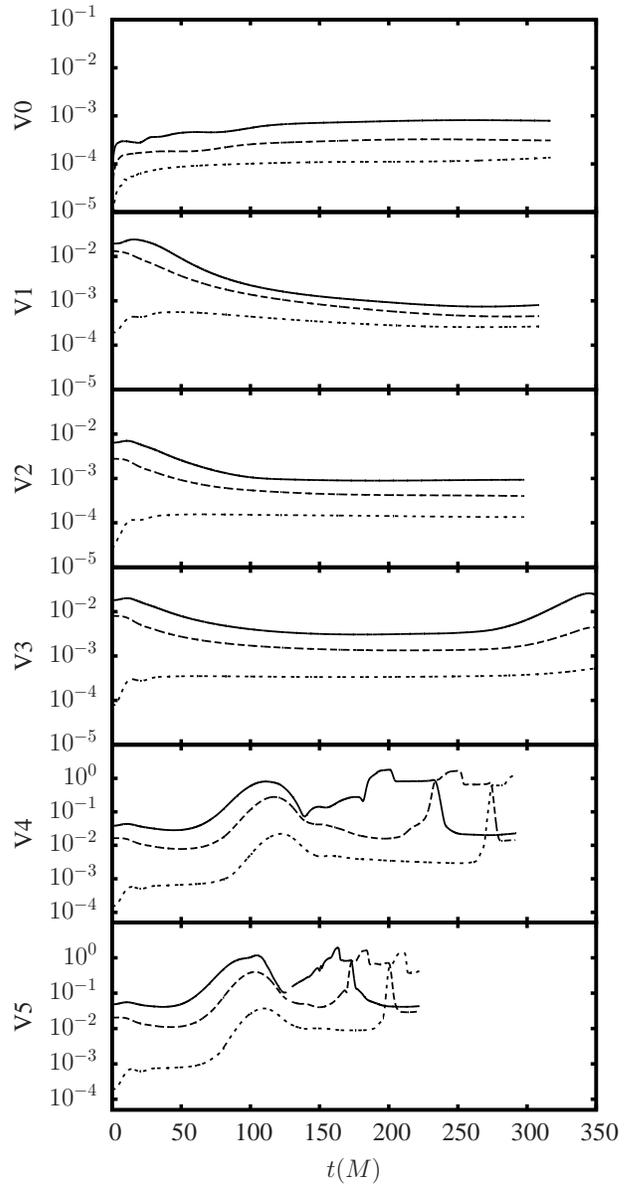}
\caption{L2 Norm of $\rho$ (r.h.s. Hamiltonian constraint) in concentric shells: $1\,M<r<2\,M$ (solid), $2\,M<r<4\,M$ (dashed), and $4\,M<r<8\,M$ (dotted).}
\label{fig:hcnorm}
\end{figure}

\begin{figure}[ht]
\includegraphics[width=0.45\textwidth]{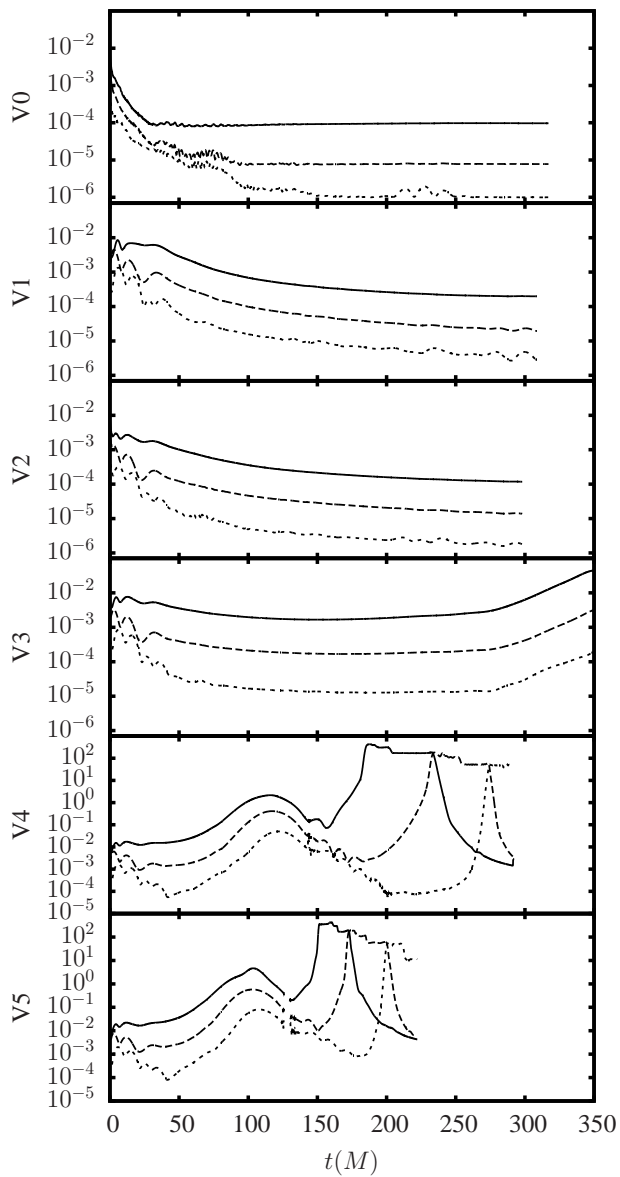}
\caption{L2 Norm of $j^i$ (r.h.s. momentum constraint) in concentric shells: $1\,M<r<2\,M$ (solid), $2\,M<r<4\,M$ (dashed), and $4\,M<r<8\,M$ (dotted).}
\label{fig:mcnorm}
\end{figure}

\begin{figure}[ht]
\includegraphics[width=0.45\textwidth]{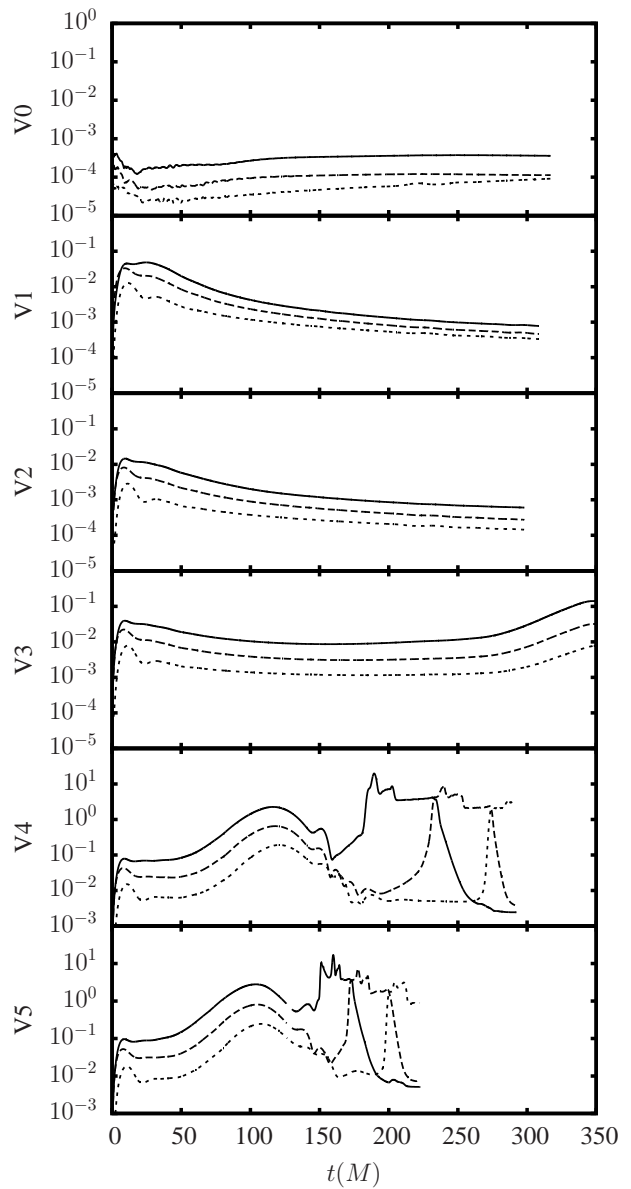}
\caption{L2 Norm of the connection constraint $\mathcal{G}^a$ in concentric shells: $1\,M<r<2\,M$ (solid), $2\,M<r<4\,M$ (dashed), and $4\,M<r<8\,M$ (dotted).}
\label{fig:vecGNorm}
\end{figure}

To understand the late behavior and, in particular, to get clues about the instability of the super-extremal \bh{s}, we calculated the L2 norms of $\rho$, $j^i$ and ${\cal G}^i $
within concentric shells $1\,M<r<2\,M$, $2\,M<r<4\,M$, and $4\,M<r<8\,M$ as a function of evolution time.
Figures~\ref{fig:hcnorm}, \ref{fig:mcnorm} and \ref{fig:vecGNorm} show these L2 norms.
First note that after $t \sim 100\,M$ the L2 norms of $\rho$, $j^i$ and ${\cal G}^i $ for the V0, V1 and V2 cases
reach comparable levels. This signals that the V1 and V2 systems are settling down to a vacuum \bh{.}  Given that the \bh{s} in V1 and V2 accreted
matter, their final mass and spins will be different from those of V0.

Case V3 shows a different behavior. Initially, $\rho$, $j^i$ and ${\cal G}^i $ drop and stabilize as with V1 and V2 as a consequence of matter having been accreted by the \bh{.}
The approximately constant values are, however, larger than the corresponding values
in V1 and V2. Beyond, $t \sim 250\,M$,  $\rho$, $j^i$ and ${\cal G}^i $ begin to grow as the \bh{} approaches its maximum $\zeta$ value at $t \sim 300\,M$ (see Fig.~\ref{fig:Zeta}).

Cases V4 and V5, those that led to super-extremal \bh{s}, show more complex behavior. Initially, the slight drop of $\rho$, $j^i$ and ${\cal G}^i $ in the outer two shells
gives a hint of the undergoing accretion with the inner shell remaining approximately constant. Around the time the \bh{} surpasses extremality,  $t \sim 65\,M$ for V4 and $t \sim 45\,M$ for V5, 
 $\rho$, $j^i$ and ${\cal G}^i $ start growing appreciably. This smooth growth continues until  $t \sim 120\,M$ for V4 and $t \sim 100\,M$ for V5. During this period,
$\zeta$ reaches a maximum value and proceeds to decrease. The decrease in $\zeta$ is eventually accompanied with a decrease in $\rho$, $j^i$ and ${\cal G}^i $.
At around $t \sim 150\,M$ for V4 and $t\sim 130\,M$ for V5, an ejection burst is triggered deep in the interior of the \bh{.} Evidence of this ejected burst is the delay at which the burst emerges in each of the shells. That is, in
Figures~\ref{fig:hcnorm}, \ref{fig:mcnorm} and \ref{fig:vecGNorm} the burst shows in the solid line first, dash line next and dotted line last.  

The L2 norms in Figures~\ref{fig:hcnorm}, \ref{fig:mcnorm} and \ref{fig:vecGNorm} do not provide a good sense of the details of the dynamics of
$\rho$, $j^i$ and ${\cal G}^i $. In Fig.~\ref{fig:rho2dsnaps} we display the temporal evolution of spatial features. We show snapshots of $\rho$ for the V4 case in the equatorial plane (left) and axis plane (right) at times 
 $t \simeq 50\,M,\,100\,M,\,150\,M,$ and $200\,M$ from top to bottom.
 The grayscale from top to bottom is such that [white:black] =  $[0:-0.003],\,[0.08:-0.08],\,[0.02:-0.02]$, and $[0.008:-0.008]$ respectively.  The panels are $5\,M$ across.
The bottom left two panels clearly show a burst of negative energy. A similar burst is found in $j^i$, which explains the smaller value of $\zeta$ during a brief period at $t\sim 150\,M$ when the \ahz{{} is again located. 
 A detailed investigation of the late behavior of these super-extremal cases will be the subject of subsequent study.

\begin{figure}[ht]
\includegraphics[width=0.45\textwidth]{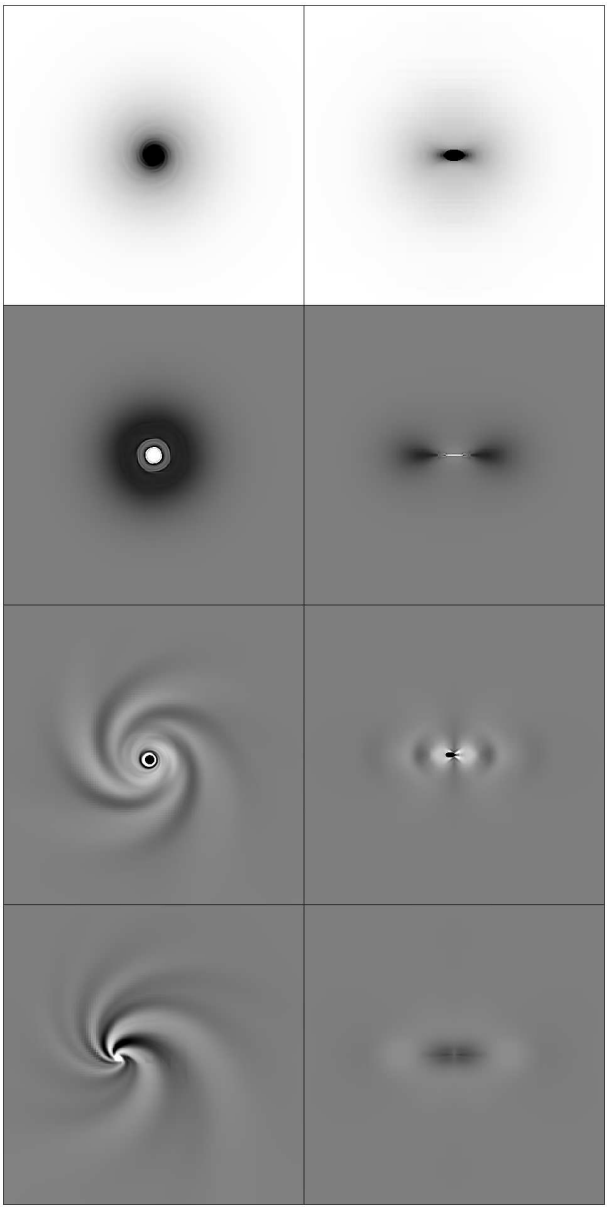}
\caption{Snapshots of $\rho$ of the V4 case in the equatorial-plane (left) and axis-plane (right) at times 
  $t \simeq 50\,M,\,100\,M,\,150\,M,$ and $200\,M$. The grayscale from top to bottom is such that [white:black] =  
  $[0:-0.003],\,[0.08:-0.08],\,[0.02:-0.02]$, and $[0.008:-0.008]$ respectively.  The regions, all centered about
  the coordinate origin, are $5\,M$ across.}
\label{fig:rho2dsnaps}
\end{figure}

\section{Conclusions}
\label{sec:end}

We have carried out a series of numerical experiments showing accreting \bh{s} that violate the
extremal spin condition $\zeta \le 1$. The experiments consisted of a \bh{,} modeled by a spinning puncture, surrounded by a spherically-symmetric
shell. We considered shells with positive energy density and negative angular momentum, but the main focus was shells with negative energy density
and positive angular momentum.
The idea behind this setup was that, as the \bh{} accretes a negative energy, positive angular momentum shell, its mass will decrease and its spin will increase,  leading to possible violations of the
$\zeta= 1$ extremality bound.
We were successful in violating the extremality bound, at least temporarily.
In agreement with the findings by Booth and Fairhurst~\cite{Booth:2007wu}, the violations were accompanied with violations of the null energy condition.
The \bh{s} that violated the $\zeta = 1$ bound were not able to sustain
super-extremality and simultaneously retain axisymmetry.

The study involved several challenges. The most significant challenge was locating the \ahz{} as the \bh{} became super-extremal.
In subsequent work, we will investigate gauge conditions that could alleviate difficulties caused by the
extreme deformations of the \bh{} horizon. If a new gauge condition is found, we will be in a better position
to investigate if violations of extremality are accompanied with the disappearance of the \ahz{.} We will also focus on the late-time behavior
to get a better understanding of the causes behind the late emission responsible for the drop of the spin parameter $\zeta$.

Our study introduced a new approach to construct non-vacuum dynamical spacetimes: the \emph{matter-without-matter} evolution framework.
Under this approach, only the evolution equations are used, with the matter source terms set to zero. Specifically, we demonstrated that the
3+1, BSSN vacuum evolution equations were capable of providing the dynamics of matter fields that are \emph{invisible} to the equations.
The price paid in \mwm{} evolutions is restrictions on the ``equations of state'' satisfied by the matter content. It is not clear whether \mwm{} evolutions
could be applied to other 3+1 formulations of the Einstein equations and produce stable and convergent simulations.

\acknowledgments
Work supported in part by NSF grants 
PHY-0914553, PHY-0855892,  PHY-0903973, PHY-0941417. Computations carried out
under TeraGrid allocation MCA08X009 and at the Texas Advanced
Computation Center, University of Texas at Austin. We thank A.~Ashtekar, S.~Fairhurst,
R.~Haas, and G.~Lovelace for their comments and suggestions.


\end{document}